\documentclass[12pt, onecolumn]{IEEEtran}

\makeatletter
\def\ps@headings{%
\def\@oddhead{\mbox{}\scriptsize\rightmark \hfil \thepage}%
\def\@evenhead{\scriptsize\thepage \hfil \leftmark\mbox{}}%
\def\@oddfoot{}%
\def\@evenfoot{}}
\makeatother 

\usepackage{amsfonts}
\usepackage{amssymb}
\usepackage{graphicx}
\usepackage{amsmath}
\usepackage{amsmath, amsfonts,epsfig, multirow, floatflt}
\usepackage{epsfig,graphics,graphicx,latexsym,amsfonts,amssymb,amsmath,verbatim,slashbox}
\usepackage{subfigure}
\usepackage{cite}
\usepackage{color}
\usepackage{soul}
\usepackage{slashbox}
\usepackage{algorithm}
\usepackage{algpseudocode}
\usepackage{multirow}

\graphicspath{{figures/}}

\begin{document}
\baselineskip 24pt
\parskip 9pt
\thispagestyle{empty}

\setcounter{page}{1}

\begin{center}
\vspace*{0mm}

{\LARGE \bf  Energy-Efficient Localization and Tracking of Mobile Devices in Wireless Sensor Networks} \vspace*{10mm}

{\normalsize Kan Zheng$^\ast$ {\it Senior Member, IEEE}, ~Hang Li$^\ast$, Lei Lei$^\dag$, Wei Xiang$^\ddag$ {\it Senior Member, IEEE},  Jian Qiao$^\natural$ and
Xuemin (Sherman) Shen$^\natural$ {\it Fellow,~IEEE}\\

\vspace{0.3cm}

\small  $^\ast$  Key Lab of Universal Wireless Communications, Ministry of Education\\
Beijing University of Posts\&Telecommunications\\
Beijing, China, 100088 \\

$^\dag$  State Key Laboratory of Rail Traffic Control \& Safety \\
Beijing Jiaotong University\\
Beijing, China, 100044\\

$^\ddag$ School of Mechanical and Electrical Engineering\\
University of Southern Queensland\\
Toowoomba, QLD 4350, Australia\\

$^ \natural$  Department of Electrical and Computer Engineering\\
University of Waterloo, Waterloo, Ontario\\
Canada~~N2L 3G1.

 }
\end{center}

\vspace*{10mm}

\begin{center} {\bf Abstract}
\end{center}
  Wireless sensor networks (WSNs) are effective for locating and tracking people and objects in various industrial environments. Since energy consumption is critical to prolonging the lifespan of WSNs, we propose an \emph{\textbf{e}nergy-efficient \textbf{LO}calization and \textbf{T}racking} (eLOT) system, using low-cost and portable hardware to enable highly accurate tracking of targets. Various fingerprint-based approaches for localization and tracking are implemented in eLOT. In order to achieve high energy efficiency, a network-level scheme coordinating collision and interference is proposed. On the other hand, based on the location information, mobile devices in eLOT can quickly associate with the specific channel in a given area, while saving energy through avoiding unnecessary transmission. Finally, a platform based on TI CC2530 and the Linux operating system is built to demonstrate the effectiveness of our proposed scheme in terms of localization accuracy and energy efficiency.
\begin{flushleft}
\end{flushleft}
\newpage
\baselineskip 24pt
\parskip 9pt

\section{Introduction}
\label{SEC_INTRO}

Location-based wireless networks are considered as one of the main technological innovations in current industrial services, which can provide high reliability and efficiency through accurately locating and tracking people and objects. For example, there are numerous applications of location-based services (LBSs) in hospitals and retail outlets, which help staff and administrators better deliver care and manage costs~\cite{ref1}\cite{ref2}. The ability of tracking the location of a subject in real time gives human operators the ability to effectively manage situations, tackle safety problems, increase efficiency, and thereby reduce costs while improving outcomes. \par

Traditional localization and tracking techniques such as GPS, cellular and Wi-Fi do not work well in many scenarios, such as high rises, underground, or disaster zones where signals from mobile infrastructure or satellites cannot be received. Neither their accuracy nor the physical size meets the demand of recent industrial applications, which aim to be highly precise in all environments even with devices of tiny sizes. Meanwhile, wireless sensor networks (WSNs) have gained much attention recently, and been in widespread use in various industrial applications including LBSs~\cite{Cenedese_TVT_10}\cite{Zhou_TVT_10}. Compared with other wireless technologies, the WSN is not only of low power consumption and
complexity, but also supports a great number of nodes in a wide coverage area. Therefore, WSN can be applied to both indoor and outdoor positioning in areas of interest. A large number of studies on localization in WSNs have been reported in the literature in recent years. \par

Usually a mobile node associated to an object in a WSN is a small device powered by battery with a limited energy budget. The battery is often disposable and inconvenient for replacement. Meanwhile, it is expected to work normally for a long enough lifetime, e.g., several months or even years. Therefore, energy-efficient techniques to reduce energy consumption are essential for wireless positioning and tracking systems. \par


In this paper, we develop an \emph{\textbf{e}nergy-efficient \textbf{LO}calization and \textbf{T}racking} (eLOT) system, which uses low-cost, portable hardware to enable highly accurate tracking of targets. In order to provide ubiquitous services both indoor and outdoor, the fingerprint localization and tracking approach with the Adaptive Weighted $K$-Nearest Neighbour (AWKNN) is proposed and implemented in eLOT.
Different from the weighted $K$-nearest neighbor (WKNN), the number of reference locations is adjustable in the AWKNN according to the surrounding environment in order to improve localization accuracy. In the eLOT system, battery recharging is assumed to be available at anchor nodes so that much more attention is paid to reducing the energy consumption at mobile nodes. Multi-radio modules may be installed at anchor nodes. One radio channel is reserved only for the wireless backhaul transmission, while another one is used for the transmission between the anchor nodes and their serving mobile nodes. By this way, interference and collision can be reduced or eliminated through channel allocation and access schemes specially designed for the eLOT system. Then, avoiding unnecessary transmission can save the energy consumption of mobile nodes. Moreover, an adaptive sounding scheme is applied at the mobile node, when it moves around the area in study. The proposed schemes can achieve an elegant tradeoff between energy efficiency and localization accuracy. \par

Moreover, we implement a demonstration platform based on TI CC2530 chips for indoor and outdoor positioning, and conduct extensive experimental studies in practical environments to verify our proposed schemes. Results show that our system attain good performances in the sense of both positioning accuracy and energy consumption. \par

The main contributions of our work are summarized as follows:
\begin{itemize}
\item An AWKNN algorithm is proposed and implemented in the eLOT system. In AWKNN, the number of reference locations can be adaptively changed during the localization process in order to achieve the high localization accuracy;
\item From the network viewpoint, channel allocation and access schemes are designed for improving energy efficiency without the loss of localization accuracy. Meanwhile, an adaptive sounding scheme is also used in the mobile node to further reduce energy consumption; and
\item Field trials are conducted with a hardware platform to demonstrate the effectiveness of the designs of eLOT.
\end{itemize}
\par

The remainder of this paper is structured as follows. Section \ref{sec_related} surveys the state-of-art. Section \ref{sec_system} presents an overview of the proposed eLOT system. In Section \ref{sec_ee_networking},  the energy-efficient networking schemes for eLOT is described.  Then, Section \ref{sec_ee_localization} presents the energy-efficient localization and tracking schemes. Sections~\ref{sec_Implementation} details the hardware and software implementation of an eLOT platform. Section~\ref{sec_Results} discusses the experiment
results, while Section~\ref{Conclusions} concludes the paper.\par

\section{Related Work}
\label{sec_related}

Most existing localization schemes can be broadly categorized into range-free and range-based approaches. Range-free approaches use metrics such as connectivity or hop counts to landmarks, or a radio map of the environment in study, and then choose the one best matching the reference as the estimated position. Generally, there exist two types of matching schemes, i.e., deterministic and probabilistic. Deterministic matching schemes compare the scalar values of some measures, e.g., the received signal strength (RSS), between the real-time values and those stored in the database, and then select the location based upon, e.g., the $K$-Nearest Neighbour (KNN), and Weighted $K$-Nearest Neighbour (WKNN)~\cite{KNN}. In probabilistic matching schemes, the location with the highest probability in the radio map is chosen, e.g., the Minimum Mean Square Error (MMSE)~\cite{MMSE}. Many experiments have demonstrated that such approaches offer good accuracy for both indoor and outdoor environments~\cite{K03}\cite{OpenMAC}. The precision and accuracy of positioning can be improved if advanced matching is applied using techniques such as neural networks~\cite{NN}. However, constructing a reliable reference such as a high quality  RSS fingerprint map is an essential part of the system for these approaches~\cite{Arya09}. Many efforts have been put forward to reduce the cost and complexity of building the reference~\cite{smart}. Different from range-free techniques, range-based approaches generally calculate the distance based on the measured characteristics of the received signals such as the RSS, direction of arrival (DOA), and so on~\cite{GZ08}. Location accuracy depends heavily on the distance discrepancy between the measured and actual distances, which is inevitable due to radio propagation~\cite{radioerror}. To reduce the effect of such discrepancy, a heuristic localization algorithm, called Probability Torus Localization (PTL), was developed to compute the position of a mobile node statistically~\cite{PTL}, in which a large number of
nodes and excessive additional hardware are required. Then, introducing a secondary antenna to the mobile node may be another effective way to improve localization accuracy without the use of expensive hardware~\cite{it}. Also, a cooperative network architecture is proposed for improving the
target detection and tracking~\cite{tvt_coop}~\cite{pd_coop}. In addition, the approach to deploy the anchor nodes for localization is investigated in~\cite{tmc_deployment}. However, nearly all the studies in the literatures focus on localization accuracy as the only key factor for achieving promising localization solutions. \par

On the other hand, there already exists substantial research related to energy-efficient techniques for WSNs~\cite{EEsurvey}\cite{Cheng13} \cite{Shu15}. The methods in relation to energy savings in the WSN can be roughly classified into two types, i.e., device level and network level. The former pays attention to nodes including hardware and configuration, whereas the latter focuses mainly on the choice of communication methods and protocols for minimizing energy consumption. However, these methods deal with traditional sensor tracking problems in WSNs, whose characteristics are far different from those of location-based applications for industrial uses with wireless infrastructure~\cite{EEtrackingsurvey}. For example, there are not too many mobile nodes in a serving area, where anchor nodes can be deployed and always powered such as in buildings. \par

The straightforward way to improve the energy efficiency of a localization system is to use low-power radio node instead of the WiFi module~\cite{tmc4}~\cite{ZiFIND}. Moreover, the power consumption of a WSN node can be minimized either by reducing the hardware power assumption or by decreasing the hardware active time at the protocol level~\cite{ville}~\cite{tmc3}. Also, context information can be exploited to reduce the frequency of updating location information to conserve energy while maintaining certain degree of accuracy~\cite{coal}. However, the performance of such schemes depends highly on the accuracy and timeliness of the context information, which is difficult to obtain in practical systems. Meanwhile, theoretical work on the tradeoff between the energy required for localization and the resulting accuracy of localization is also carried out. In~\cite{santosh}, both the centralized and distributed implementations of the range-based localization schemes are discussed, and a linear relationship between energy efficiency and localization accuracy is established under some simplified assumptions. A measure to characterize the energy efficiency of localization algorithms in wireless networks is also presented in ~\cite{dominik}.\par

Although energy-efficient localization has been active research subbject, there is still little work on designing localization systems with both high energy efficiency and good localization accuracy. Furthermore, the location information of mobile nodes has not been well exploited to assist in the energy-efficient design. Therefore, the proposed eLOT system can address these aforementioned issues.\par

\section{System Overview}
\label{sec_system}

\begin{figure}
\centering
\includegraphics[width=2.5 in, angle=270]{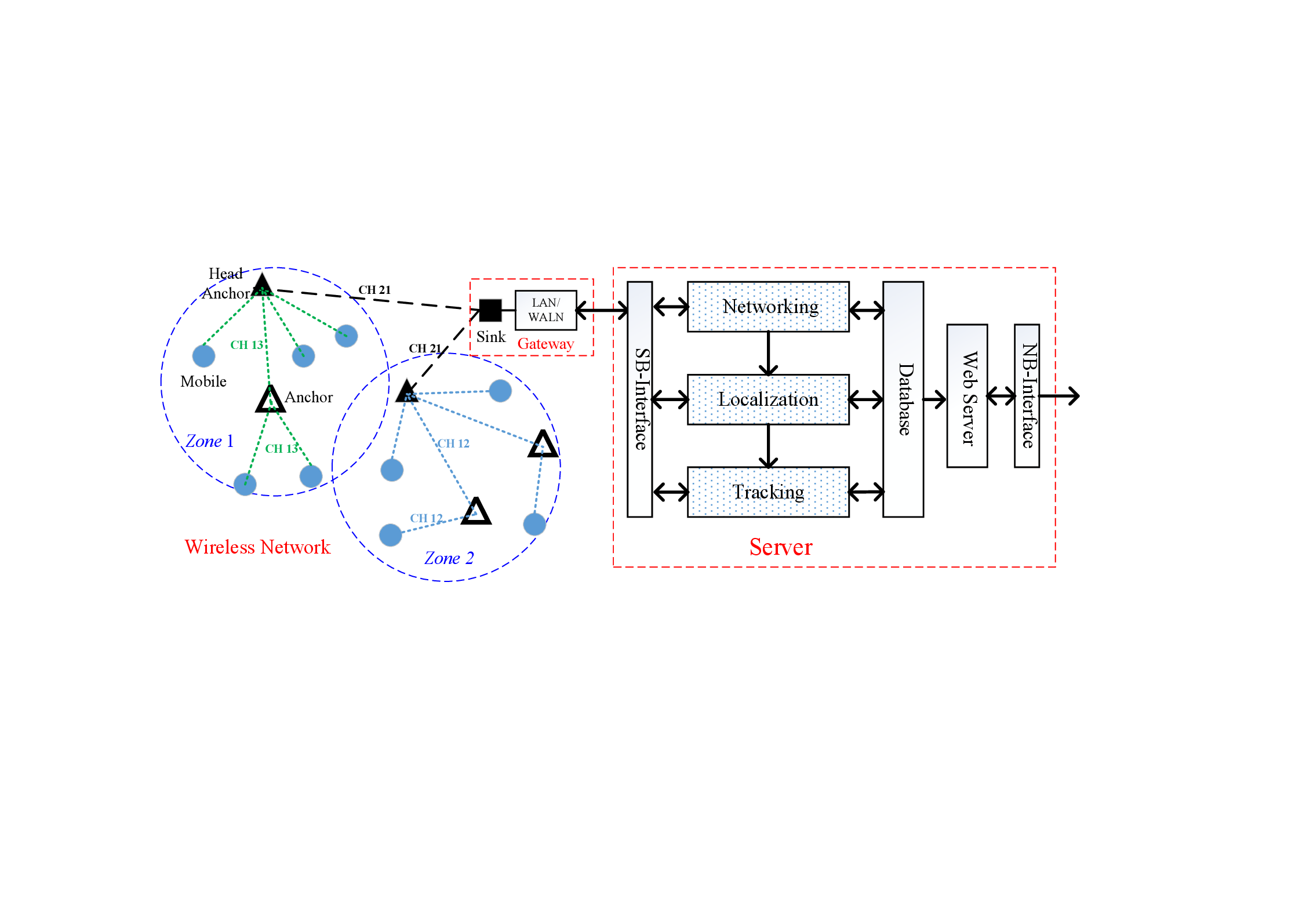}
\caption{Illustration on the proposed eLOT system.}
\label{fig_eLOT}
\end{figure}

As shown in Fig.~\ref{fig_eLOT}, the eLOT system consists of a wireless network, a gateway and a server. With the aid of the gateway, the information collected by the wireless network can be forwarded to the server, which is responsible for positioning and tracking. These subsystems will be detailed in the following.

\subsection{Wireless Network}
Since low power consumption is one of main focuses, the wireless network with ZigBee based on the IEEE 802.15.4 standard is chosen for the eLOT system. A ZigBee device can transmit data over a long distance by passing data through a mesh network of intermediate devices to reach other distant devices. There are three kinds of nodes existing in wireless networks,
\begin{itemize}
  \item {\emph{Mobile node (MN):}} \par
Mobile nodes (MNs), as position and tracking targets, are equipped with a
single radio. They send beacon signals periodically to surrounding anchor nodes when moving around their serving area. MNs may sleep between beacons, thus extending their battery life. The duty cycle can be adaptively configured according to application requirements.
   \par
  \item {\emph{Anchor node (AN):}}\par
Anchor nodes (ANs) are deployed uniformly in the serving area. The location information of an anchor node is stored at the server. In order to grant mobile nodes timely access to the network at anytime, anchor nodes do not fall sleep. This is easy in the indoor environment, e.g., plugging in a wall power outlet. On the other hand, the solar panel can be used for recharging anchor nodes deployed outdoor. The entire serving area is divided into several zones. The ANs in the same zone form a cluster, in which one of them is chosen as the head anchor node (HAN) according to an appropriate criterion. Clustering can improve scalability and simplify routing protocols,
and thus increase energy efficiency. Multiple radios are installed in the HAN, which can work on multiple channels simultaneously. For a common AN, a single radio is enough to maintain communications with others. The RSS data from the MNs in the zone are continuously collected by the ANs and forwarded to a sink node with single-hope or multi-hop transmission.
    \item {\emph{Sink node (SN):}}\par
There exists only a single sink node (SN) in a wireless network. The SN collects data from all the HANs deployed in the field, and sends data to the server. Usually, it is installed together with the gateway for the convenience of deployment.
\end{itemize}

\subsection{Gateway}

The gateway in eLOT is a device that can support both data transmission for both Internet access via a WLAN or LAN interface, and remote MNs via the Zigbee network. This can be implemented on either an ARM board or a computer. A gateway facilitates communications between the nodes and server. \par

\subsection{Server}
All RSS information is collected and stored at the server, which is responsible for the localization and tracking process. In order to fulfil its duties, the server is equipped with several key processing modules, e.g., the networking, localization and tracking modules. Meanwhile, the RSS database is constructed and maintained, which works together with the other modules. All localization and tracking functions are implemented at the server, while the mobile nodes only report the RSS data. With the centralized control, one can conveniently upgrade the localization and tracking scheme in eLOT if needed.\par


\section{Energy-efficient Networking Schemes}
\label{sec_ee_networking}

To enable high energy-efficient wireless networks, one needs to consider from both sensor network and device aspects. In eLOT, channel allocation schemes are employed to eliminate the interference between zones, while coordinated backoff schemes are used for alleviating the intra-zone interference. On the other hand, based on location information, mobile nodes in eLOT can quickly associate with or switch to a zone-specific channel, while saving energy by reducing unnecessary transmissions. \par

\subsection{Channel Allocation Scheme}

In the eLOT system, there are usually three types of links as illustrated in Fig.~\ref{fig_eLOT}, namely, the MN-to-AN, AN-to-HAN, and
HAN-to-HAN links. Since the wireless backhaul link, i.e., the HAN-to-HAN link, carries large amounts of RSS data, its fast and reliable transmission is essential for achieving good system performance. Thanks to the multi-radio configuration at the HAN, it is possible for the eLOT system to allocate a dedicated channel to this link, e.g., \emph{Channel} 21, to guarantee the QoS performance of the backhaul transmission. Another two types of links, i.e., the MN-to-AN and AN-to-HAN links, share the rest of available channels. Then,
collision and interference become challenging issues. Therefore, it is essential to first design efficient channel allocation schemes in such
networks. \par

As mentioned in Section \ref{sec_system}, the serving area is geographically divided into several zones, each of which is allocated one of the available channels. When an MN roams into a given zone, it is then allowed to utilize the channel allocated to the new zone. Different channels can be used in neighboring zones, e.g., \emph{Channel} 13 in \emph{Zone} 1 and \emph{Channel} 12 in \emph{Zone} 2 as shown in Fig.~\ref{fig_eLOT}. The same channel can be reused by different zones, only when these zones are separated by a minimum distance.
By this way, the collision and interference between the MN-to-AN links in the neighbouring zones can be completely eliminated.\par

\begin{figure}
\centering
\includegraphics[width=3 in, angle=270]{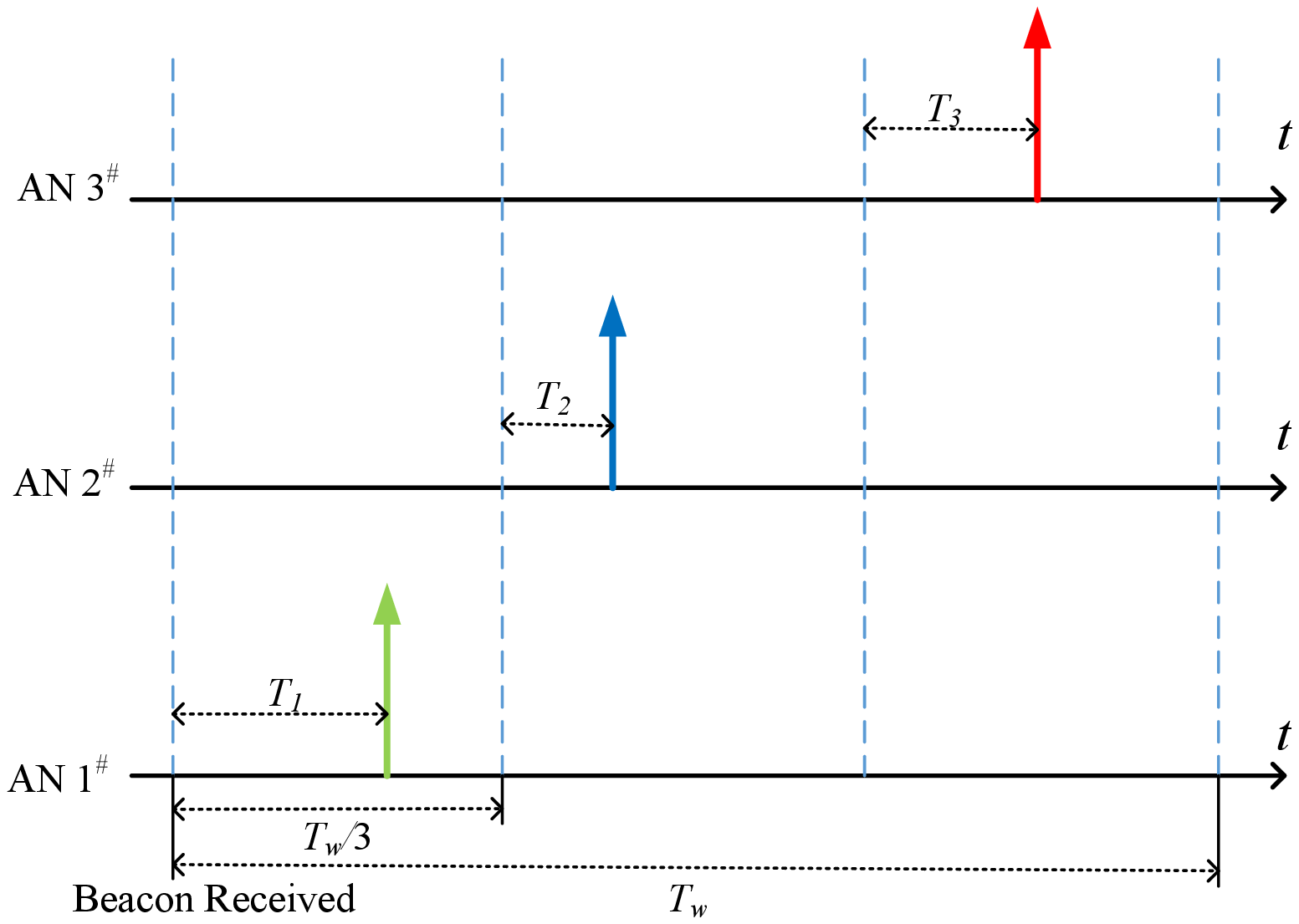}
\caption{An example of the coordinated backoff scheme ($N_a$=3).}
\label{fig_backoff}
\end{figure}

Moreover, in order to reduce the possible interference between the AN-to-HAN links in the same zone, a coordinated backoff scheme in the time domain is proposed for the eLOT system. When an MN sends a beacon signal, the nearby $N_a$ ANs in the same zone receive it almost instantaneously due to geographical proximity. If each AN forwards its received RSS information immediately, serious collision may occur. So, it is better for the ANs to wait for a while, i.e., the backoff period, before transmission. Firstly, each AN generates a random number using a common pseudo-random number generator,
which is used for calculating a random period, i.e., $T_i$, $1 \leq i \leq N_a$. On the other hand, the time window, i.e., $T_w$, is pre-defined by the system and equally divided into several fixed time slots. Then, the $i$th AN starts transmitting signals at the moment of $T_i+iT_w/N_a$. Fig.~\ref{fig_backoff} illustrates an example of the coordinated backoff scheme with $N_a=3$ ANs. \par

In the above centralized schemes, channel and time slot allocation for the entire network is managed by the networking module at the server, which is responsible for coordinating all the channels. \par

\subsection{Location-based Channel Access Scheme}

When there are a large number of MNs in the system, serious collision between beacon signals may take place, if all the MNs work on the same channel. As mentioned above, different channels to the MNs are assigned in different zones, and the timely switch is required when the MNs roam cross the border. The details of a location-based channel access and switch scheme are discussed as follows.\par

\begin{itemize}
  \item {\emph{Initialization:}} \par
All the HANs can be configured through a dedicated channel by the server, e.g., \emph{Channel} 21. Different zones correspond to different channels. As a result, in a given zone, the HAN sets one of its radio to the channel according to the commands received from the dedicated channel. \par
   \par
  \item {\emph{Channel Access:}}\par
After entering its serving area, an MN scans all the channels for AN discovery, e.g., from \emph{Channel} 11 to \emph{Channel} 26 except \emph{Channel} 21. If the MN finds an active channel in a given zone and associate itself successfully,  the beacon signal is transmitted either periodically or driven by events.
    \item {\emph{Channel Switching:}}\par
All the RSS data are forwarded to the server through the SN. The channel switching function in the networking module keeps checking the number of the ANs successfully receiving the beacon signal from a given MN, and the corresponding RSS values of the received signal. If the number of the ANs with a strong enough RSS is less than a specific threshold, the decision of channel switching has to be made. Assume that the position history information of this MN is available in the database, a candidate destination channel will be selected according to the pre-defined channel allocation scheme in the neighboring zone the MN may move in. Then, the identification of the candidate channel is sent through the dedicate channel to the responsible HAN. Next, the HAN transfers this information to the MN, which may re-establish a new connection if needed.\par
\end{itemize}
The detailed algorithm is presented in Algorithm~\ref{alg1}.  \par

\begin{algorithm}
\caption{ \quad Location-based Channel Access and Switch Scheme} \label{alg1}
\begin{algorithmic}
\State \textbf{Step 0: Head anchor node initialization.}

Head-Anchor-Node's up\_radio on Channel

$HAN_n.up\_radio {\rm{ = }}CH_{21}$

Head-Anchor-Node's down\_radio on the channel according to the commands received from the dedicated channel.

$HAN_n.down\_radio {\rm{ = }} CH_x$

\State \textbf{Step 1: Anchor node initialization.}

\textbf{loop} $n=1...$ for each $AN_k$

Anchor-Node's down\_radio's channel depends on area

\quad \textbf{loop} $CH_{11}$ to $CH_{26}$ except $CH_{21}$

\quad \quad \textbf{if} $CH_n$ in $Zone$, \textbf{then}

\quad \quad \quad $AN_n.down\_radio{\rm{ = }}CH_n$.

\quad \quad \textbf{end if}

\quad \textbf{end loop}

\textbf{end loop}

\State \textbf{Step 2: Mobile node: channel access.}

\textbf{loop} $k=1...$ for each $MN_k$

\quad \textbf{loop} $CH_{11}$ to $CH_{26}$ except $CH_{21}$

\quad \quad \textbf{if} Access Successful, \textbf{then}

\quad \quad \quad \textbf{break}

\quad \quad \textbf{end if}

\quad \textbf{end loop}

\textbf{end loop}

\State \textbf{Step 3: Mobile node: channel switch.}

\textbf{loop} $k=1...$ for each $MN_k$

\quad \textbf{if} $MN_k$ receive data, \textbf{then}

\quad \quad \textbf{if} $MN_k$ on the border of an area, \textbf{then}

\quad \quad $MN_k.channel{\rm{ = }}MN_k.receive\_data\left[channel\right]$.

\quad \textbf{else}

\quad \quad $MN_k$ sends data

\quad \textbf{end if}

\textbf{end loop}

\end{algorithmic}
\end{algorithm}


\section{Localization and tracking schemes}
\label{sec_ee_localization}
This section discusses the techniques for localization and tracking schemes applied in eLOT.  The process consists of three steps, i.e., the offline phase for radio map generation, online localization phase and tracking phase. In the offline phase, a fingerprint database of the area under study is constructed which includes all necessary radio parameters such as the AN identifications (IDs), the RSS values and so on. Then, in the online phase, an MN sends the beacon signal to the nearby ANs, which can estimate the MN's location through matching the received RSS values with those fingerprint entries in the database. This can be achieved by implementing various localization algorithms. in the last phase, the MN is tracked when it moves in the serving area. \par

Without loss of generality, the two-dimensional space $\mathbb{S}$ is used to represent the serving area in this paper. Assume that there are $N$ ANs deployed for localization. Thus, corresponding to one given location, an element in this space is a $N$-dimensional vector whose entries are the RSS samples received at the ANs, i.e., $\mathbf{R}=[r_{1},r_{2}, \cdots, r_{N}]^{T}$. Therefore, the problem is to find the location $\mathbf{S}=(x,y) \in \mathbb{S}$ that maximizes the probability of $P(\mathbf{S} | \mathbf{R}) $ given an RSS vector $\mathbf{R}$.\par

\subsection{Offline Radio Map Generation}
In the offline phase, the ANs deployed in the reference points measure the RSS of the received signal from the test MN moving around. The entire area under study is logically divided into $L$ square regions. The choice of the grid spacing has a significant effect on the performance of localization. A large grid spacing results in low localization accuracy, while a small one leads to an excessive amount of time in collecting an adequate number of RSS samples for the database. At each grid, a set of continuous samples are measured and averaged to generate a raw fingerprint. Moreover, due to the unstable wireless signal propagation effect, noise-corrupted samples may lead to an unreliable fingerprint, which needs to be detected and filtered out. So, the samples with the maximum bias are first removed. Then, the fingerprint is calculated by averaging the remaining samples and stored in the database. Although this method is not necessarily the optimum approach, it is employed by eLOT due to its simplicity.\par

The database consists of a set of fingerprints along with their corresponding grid information, i.e.,
\begin{eqnarray}
\label{fingerprint}
\mathbf{\Omega}=\{(\bar{\mathbf{S}}_{i},\bar{\mathbf{R}}_{i})|i= 1,2, \cdots, L \},
\end{eqnarray}
\noindent where $\mathbf{\bar{S}}_{i}=(x_{i},y_{i})$ is the test location of the $i$th grid with the coordinates of $x_i$ and $y_i$,
and $\mathbf{\bar{R}}_{i}$ is the RSS set received at the ANs, i.e.,
\begin{eqnarray}
\label{RSSset}
\bar{\mathbf{R}}_{i}=[r_{i,1},r_{i,2}, \cdots, r_{i,L}]^{T}, i=1, 2, \cdots, L,
\end{eqnarray}
\noindent where $r_{i,j}$ is the RSS at the $j$th ANs from the test MN located in the $i$th grid.\par

\subsection{Online Localization}
In the online phase, an interested MN at location $\tilde{\mathbf{S}}$ sends the beacon signal, and its RSS is measured by the surrounding ANs as $\tilde{\mathbf{R}}$. These information is forwarded to the localization module at the server in time. Instead of giving the most probable location, the $K$ closest matches of known locations in signal space from the previously-built database are chosen for estimation in eLoT for the sake of implementation.\par

The location estimate can be expressed in a simple mathematical formulation as follows:
 \begin{eqnarray}
\label{equ:dl:rxsignalrearranged}
\hat{\mathbf{S}} =\underbrace{\left[ \begin{array}{c} w_1, w_2, \cdots,
w_L \end{array} \right]}_{
\mathbf{W}}  \underbrace{\left[
\begin{array}{cccc}
a_{1} & 0 &\cdots & 0 \\
0 & a_{2} &\cdots & 0 \\
0 & 0 &\ddots & 0 \\
0 & \cdots &0 & a_{L} \\
\end{array} \right]
}_{\mathbf{A}}
\underbrace{\left[ \begin{array}{c} \mathbf{\bar{S}}_1\\ \mathbf{\bar{S}}_2 \\ \vdots \\
\mathbf{\bar{S}}_L \end{array} \right]}_{
\mathbf{\bar{S}}}
\end{eqnarray}
 \noindent where $\bar{\mathbf{S}}=[\bar{\mathbf{S}}_{1}, \bar{\mathbf{S}}_{2},\cdots,\bar{\mathbf{S}}_{L}]^{T}$ represents the reference location, and the diagonal matrix $\mathbf{A}$ contains the selection coefficients describing which references are chosen. There are three deterministic schemes are implemented in eLOT as follows, i.e., \par

%

\subsubsection{$K$-Nearest Neighbour (KNN)}
Let $\mathbf{A}_{(i)}=[0,\cdots,a_{i},\cdots,0]$ represent the $i$th row of $\mathbf{A}$, which generates the $i$th component for position estimation. The KNN deterministic localization algorithms compute the distance between the measured $\tilde{\mathbf{R}}$ at the unknown location and the stored fingerprint $\bar{\mathbf{R}}_{i}$, i.e.,
 \begin{eqnarray}
\label{distance}
D_{i}=||\bar{\mathbf{R}}_{i}-\tilde{\mathbf{R}}||.
\end{eqnarray}
Then, the $K$ reference locations with the smallest distances are chosen, i.e.,
 \begin{equation}
a_{i}=\left\{ \begin{array}{ll}1, & \forall i \in \mathbf{\Omega} \\
0,  & \mathrm{otherwise} \\
\end{array} \right.
\label{equ:chanesti:initcompletedg}
\end{equation}
\noindent where $\mathbf{\Omega}$ is the position set with the $K$
minimum Euclidean distance. Correspondingly, the weight matrix is simply given by
 \begin{eqnarray}
\label{KNNweight}
\mathbf{W}=\frac{1}{K}\mathbf{I}
\end{eqnarray}

Therefore, the average of the $K$ locations is chosen as the estimated result, i.e.,
 \begin{eqnarray}
\label{KNN}
\hat{\mathbf{S}}=\frac{1}{K}\sum_{i=1, i \in \mathbf{\Omega}}^{L} \bar{\mathbf{S}}_{i}.
\end{eqnarray}


\subsubsection{Weighted $K$-Nearest Neighbour (WKNN)}

Different from KNN, different weights are assigned to the $K$ chosen reference locations in the WKNN algorithm. It is clear that how to choose the weights may significantly affect the localization performances. For example, the Euclidean distance between the chosen reference location and the unknown location can be taken as the corresponding weight, i.e.,
 \begin{equation}
\label{WKNNweight2}
w^{'}_{i}=\left\{ \begin{array}{ll}D_{i}^{-1}, & \forall i \in \mathbf{\Omega} \\
0,  & \mathrm{otherwise} \\
\end{array} \right.
\end{equation}
Then, these weights are normalized as
 \begin{equation}
\label{WKNNweight3}
w_{i}= \frac{w^{'}_{i}}{\sum_{j=1,}^L w^{'}_{j}}, i=1, 2, \cdots, L.
\end{equation}
Thus,
 \begin{eqnarray}
\label{WKNN}
\hat{\mathbf{S}}= \sum_{i=1, i \in \mathbf{\Omega}}^{L} w_{i} \bar{\mathbf{S}}_{i}.
\end{eqnarray}


\subsubsection{Adaptive Weighted $K$-Nearest Neighbour (AWKNN)}
Besides the weights, it is also very important to choose a proper number of reference locations, i.e., $K$, which depends not only on the surrounding wireless channel characteristics but also on geographical features of the area under consideration. Therefore, instead of choosing a fixed value, we propose to adjust the value of $K$ in accordance with the changing environments. \par

\begin{itemize}
  \item {\textbf{Step 1}:} A threshold $\Gamma(t), t=0$ for selecting the estimated distances is initialized firstly.
\item {\textbf{Step 2}:} Then, all the locations with the distance less than $\Gamma (t)$ are chosen to be the candidate reference locations, i.e.,
  \begin{equation}
a_{i}=\left\{ \begin{array}{ll}1, &  \mathrm{if} \ \ D_{i} \leq \Gamma (t) \\
0,  &\mathrm{otherwise} \\
\end{array} \right.
\label{equ:awknnw}
\end{equation}
Generally, the higher the threshold, the more reference locations are to be selected, and vice versa. Further adjustment is needed to ensure the effectiveness of the threshold setting as well as the appropriateness of the number of the reference locations.
\item {\textbf{Step 3}:} The number of selected reference locations has to be checked, i.e.,
  \begin{eqnarray}
\label{WKNNNr}
N_R=\sum_{i=1}^{L}a_{i}.
\end{eqnarray}
If $N_{R} \leq 3$, the number of reference locations is too small to produce valid localization results. Then, the threshold has to be increased by the step of $\Delta$, i.e.,
  \begin{eqnarray}
\label{thrdadjust}
\Gamma (t+1)=\Gamma (t)+\Delta.
\end{eqnarray}
Next, the procedure goes to \textbf{Step 2}. Otherwise, the following steps are executed.
\item {\textbf{Step 4}:} The distances of the $N_{R}$ selected locations are sorted in the ascending order, i.e., $\tilde{D}_{1} < \tilde{D}_{2} \cdots <\tilde{D}_{N_{R}}$. Then, the ratio between the largest and smallest distances is calculated as
     \begin{eqnarray}
\label{thrdadjust2}
\varepsilon=\tilde{D}_{N_R}/\tilde{D}_1.
\end{eqnarray}
The threshold is adjusted by
  \begin{eqnarray}
\label{thrdadjust2}
\Gamma (t+1)=\left\{ \begin{array}{ll}\Gamma (t)-\Delta, &  \mathrm{if} \ \ \varepsilon > \Theta_{L} \\
\Gamma (t)+\Delta,  &\mathrm{if} \ \ \varepsilon < \Theta_{S} \\
\Gamma (t),  &\mathrm{otherwise} \\
\end{array} \right. .
\end{eqnarray}
\noindent where $\Theta_{L}$ and $\Theta_{S}$ are the control parameters.
\end{itemize}

\textbf{Step 2} through \textbf{Step 4} are repeated as necessary until no change is needed for the threshold, which means that a suitable $K$ is found. When the iterative procedure is finished, the estimated location result can be calculated as the WKNN algorithm. \par


\subsection{Tracking}
One of the most effective ways to improve the energy efficiency of an MN is to set its radio transceiver into the sleep mode as much as possible. However, since the system has to track the movement of the MN in a timely and accurate manner, it is necessary to design a suitable tracking scheme to strike a good balance between energy efficiency and localization accuracy. \par

When an MN moves within the area under study, its radio switches on and off periodically, only sending the beacon signal when necessary. An adaptive sounding scheme is proposed to improve the energy efficiency of the MNs being tracked. When the MNs are powered on, the default value of the duty cycle for sounding is used for transmission. Then, there are two conditions, in which the configuration of the sounding cycle needs to be adjusted, i.e.,
  \begin{itemize}
  \item {\emph{Mobility event driven}:}\par
When the tracking module at the server detects the mobility variance is larger or smaller than the specified threshold, it may send the adjustment command to the MNs through the ANs. In principle, the smaller the sounding cycle is, the higher the speed is, and vice versa. However, there is no need to adjust the sounding cycle continuously or too frequently. As it is known, speed measurement errors often occur due to the varying wireless signals. As unwise adjustment may lead to energy over-consumption, it is better to select a proper duty cycle according to a range of speeds instead of a specific value. After careful field measurements, the recommended adjustment threshold is given in Table \ref{tab_dutycycle}. Moreover, only after detecting multiple results with the required speed, the adjustment command is sent to the specific MN.\par
  \item {\emph{Location event driven}:}\par
When the MN moves across the edge of the zone, its sounding cycle increases after receiving the downlink command from the server. Then, it may quickly receive the channel switching message when at the zone edge.
  \end{itemize}
The detailed algorithm of the location-aware tracking scheme is presented in Algorithm~\ref{alg2}.
\par

\begin{table}
\centering
\renewcommand{\arraystretch}{1.7}
\caption{Parameters for adjusting the duty cycle for sounding. } \label{tab_dutycycle}
\begin{tabular}{|c|c|}
\hline
 \textbf{Speed range} & \textbf{Duty cycle}  \\ \hline
$<$ 0.1 $m/s$ &	$>$ 2 $s$  \\ \hline
0.1 $\sim$ 0.5 $m/s$ &	2 $s$ \\ \hline
0.5 $\sim$ 1 $m/s$	& 1 $s$ \\ \hline
1 $\sim$ 1.5 $m/s$ &	0.5 $s$ \\ \hline
$>$ 1.5 $m/s$ 	& 0.2 $s$ \\ \hline
\end{tabular}
\end{table}

\begin{algorithm}
\caption{ \quad Location-aware Tracking Scheme } \label{alg2}
\begin{algorithmic}

\State \textbf{Step 0: Mobile-Node's duty-cycle configuration initialization.}

$MN_n.transmit\_period{\rm{ = }}1s$

$MN_n.receive\_period{\rm{ = }}3s$

\State \textbf{Step 1: Mobile-Node's transmit\_period adjustment.}

\textbf{if} $MN_n.speed\ increased$

\quad $Decrease\ MN_n.transmit\_period$

\textbf{end if}

\textbf{if} $MN_n.speed\ decreased$

\quad $Increase\ MN_n.transmit\_period$

\textbf{end if}

\State \textbf{Step 2: Mobile-Node's recieve\_period adjustment.}

\textbf{if} $MN_n.position{\rm{ = }}Edge\_of\_an\_Area$

\quad Mobile-Node is on the edge of an area

\quad $Decrease\ MN_n.receive\_period$

\textbf{end if}

\textbf{if} $MN_n.position{\rm{ = }}Center\_of\_an\_Area$

\quad Mobile-Node is in the center of an area

\quad $Increase\ MN_n.receive\_period$

\textbf{end if}

\end{algorithmic}
\end{algorithm}

\section{Implementation}
\label{sec_Implementation}
The purpose of our implementation is to demonstrate the effectiveness of the proposed schemes in terms of energy efficiency and localization accuracy. This section presents the details of both hardware and software implementations of eLOT.\par

\subsection{Hardware Platform}
CC2530 is chosen to be the core chipset to implement the wireless communications module on a self-designed PCB. It is TI's second generation ZigBee/ IEEE 802.15.4 RF System-on-Chip (SoC) for the 2.4 GHz unlicensed ISM band~\cite{2530}. We only use 16 channels at 2.4 GHz in eLOT, i.e., from \emph{Channel} 11 to \emph{Channel} 26, as defined in IEEE 802.15.4~\cite{802154}. \par

This chip enables industrial grade applications by offering good selectivity/co-existence, excellent link budget, and low voltage operation. It enables robust network nodes to be built with very low total bill-of-material costs. CC2530 combines the excellent performance of a leading RF transceiver with an industry-standard enhanced 8051 MCU, in-system programmable flash memory, 8-KB RAM, and many other powerful features. CC2530 can be equipped with TI standard compatibility or a proprietary network protocol stack, e.g., Z-Stack and SimpliciTI, to streamline the development process. \par

However, the output power of the RF transmitter of CC2530 is only 4.5 dBm with a receiver sensitivity of -97 dBm. Thus it is unable to satisfy the requirements of localization applications. As a result, CC2591 is employed in our module as a range extender for the 2.4-GHz RF transceivers, which increases the link budget by providing a power amplifier (PA) for higher output power and a low noise amplifier (LNA) for improved receiver sensitivity. Then, the maximum transmit power is increased to 20 dBm. When using CC2530 with CC2591, duty cycling or back-off is needed for the highest IEEE 802.15.4 channel, i.e., \emph{Channel 26}, so as to comply with the FCC with regards to the maximum recommended output power. \par

\begin{figure}
\centering
\subfigure[Structure diagram. ]{\includegraphics[width=0.4\textwidth, angle=270]{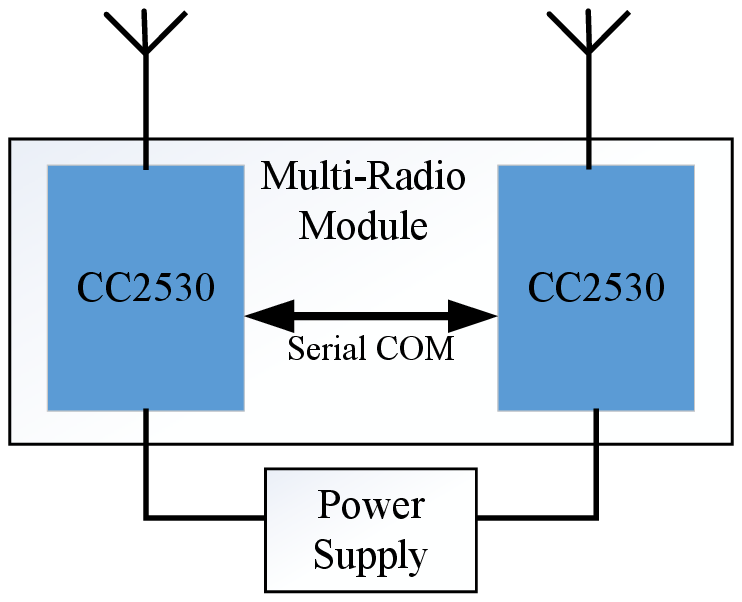}}
\subfigure[Photo.]{\includegraphics[width=0.4\textwidth, angle=270]{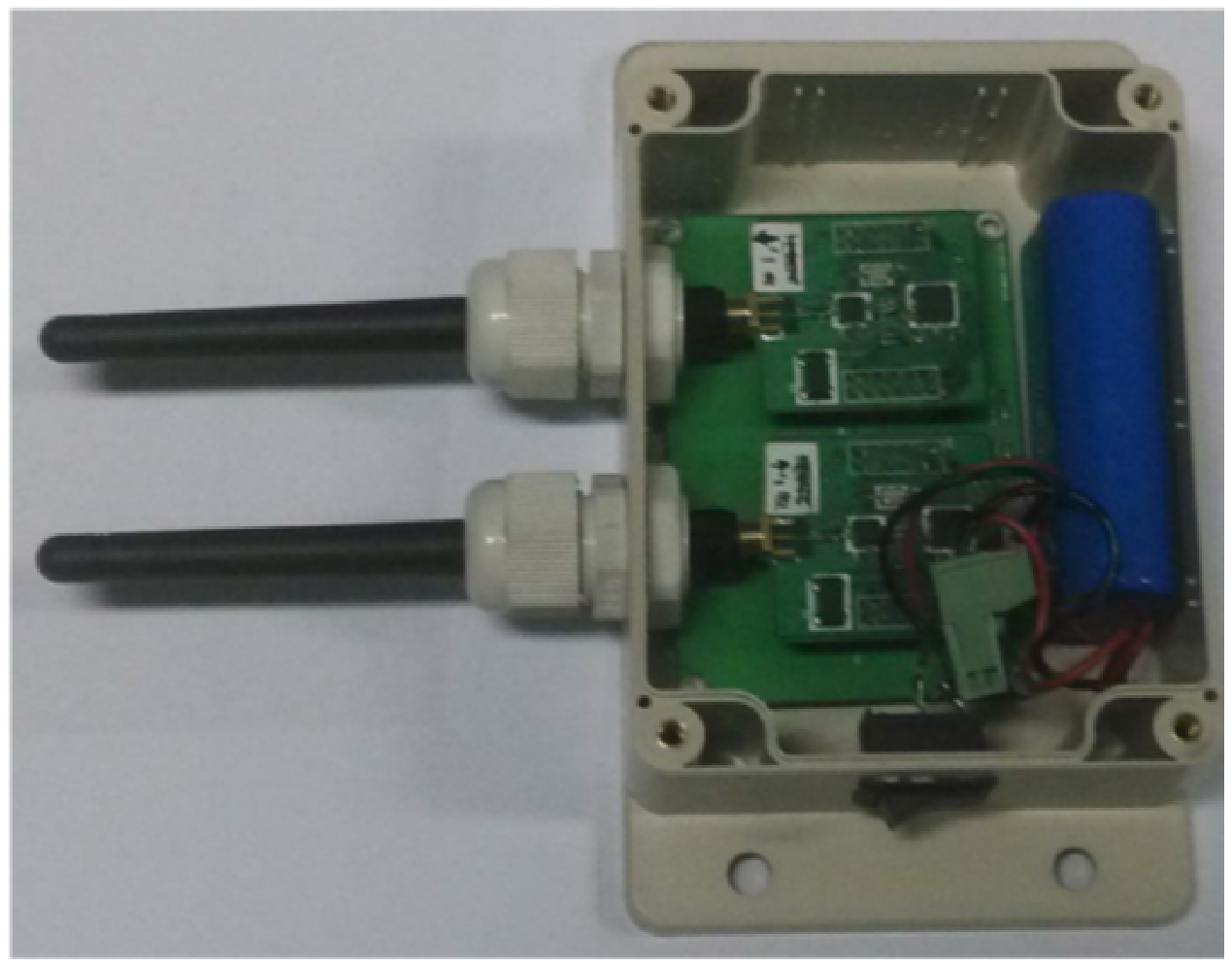}}
 \caption{Illustration of the multi-radio wireless communications module.}
 \label{multiradio}
\end{figure}

In eLOT, most nodes are installed with a single-radio wireless communications module, e.g., the mobile nodes, sink node and common anchor nodes. For fast and efficient data forwarding, the head anchor nodes are equipped with a multi-radio wireless communication module, which is implemented by connecting two single-radio modules through a serial port as shown in Fig.~\ref{multiradio}.\par

\subsection{Software Platform}
All the software modules are programmed and run in the Linux operation systems. Besides the key modules such as the networking, localization and track modules, there are also other operation and maintenance modules in eLOT, which facilitate control and demonstration.

\subsubsection{Interface Module}
This module consists of two parts, i.e., the South-Bound Interface and North-Bound Interface. Due to the centralized control mechanism in eLOT, the server needs to frequently exchange information with the underlying infrastructure via the South-Bound Interface, including various RSS and network status data. As a result, the socket protocol is used to implement the South-Bound Interface to ensure effective and reliable communications. On the other hand, the North-Bound Interface provides communications between the database and the browsers accessed by remote users, which is implemented by the WebSocket protocol. In the WebSocket protocol, messages are allowed to be passed back and forth maintaining an open connection. Hence, fast two-way (bi-directional) communications are enabled between a browser and the server.

\subsubsection{Database Module}
Huge amounts of data are to be stored and processed in eLOT. The open source MySQL database is chosen due to its excellent reliability and scalability. There are mainly two types of data established and operated in the database detailed as follows:
  \begin{itemize}
  \item {\emph{Offline Data}:}\par
Typically, two tables are maintained in the database as the radio fingerprint map. The first table records all the IDs and coordinates of the reference locations. The corresponding RSS measurements are stored in the second table.
\item {\emph{Online Data}:}\par
When the localization and tracking modules process the collected real-time data, they produce the position estimates of the MNs as well. These results are saved in another table, which contains the MN ID, the coordinates of the estimate, the timestamp, and so on.
 \end{itemize}

Instead of using a database table, a hash table is created and loaded in the memory of the server to expedite the I/O processes. This table is used to save the real-time data, collected from the ANs. The main fields in the table include the MN ID, RSS value, Timestamp, the ID of the reference AN. Each MN has its own linked list in the hash table. The hash value for a linked list of an MN is generated according to the MN ID, which determines its position in the hash table. Since the memory for a linked list is dynamically allocated, it may move the data due to an improper assigned memory size. Thus, by this way, it is convenient to store all the MNs¡¯ data efficiently. \par

\subsubsection{Web Server Module}
\begin{figure}
\centering
\includegraphics[width=0.5\textwidth, angle=270]{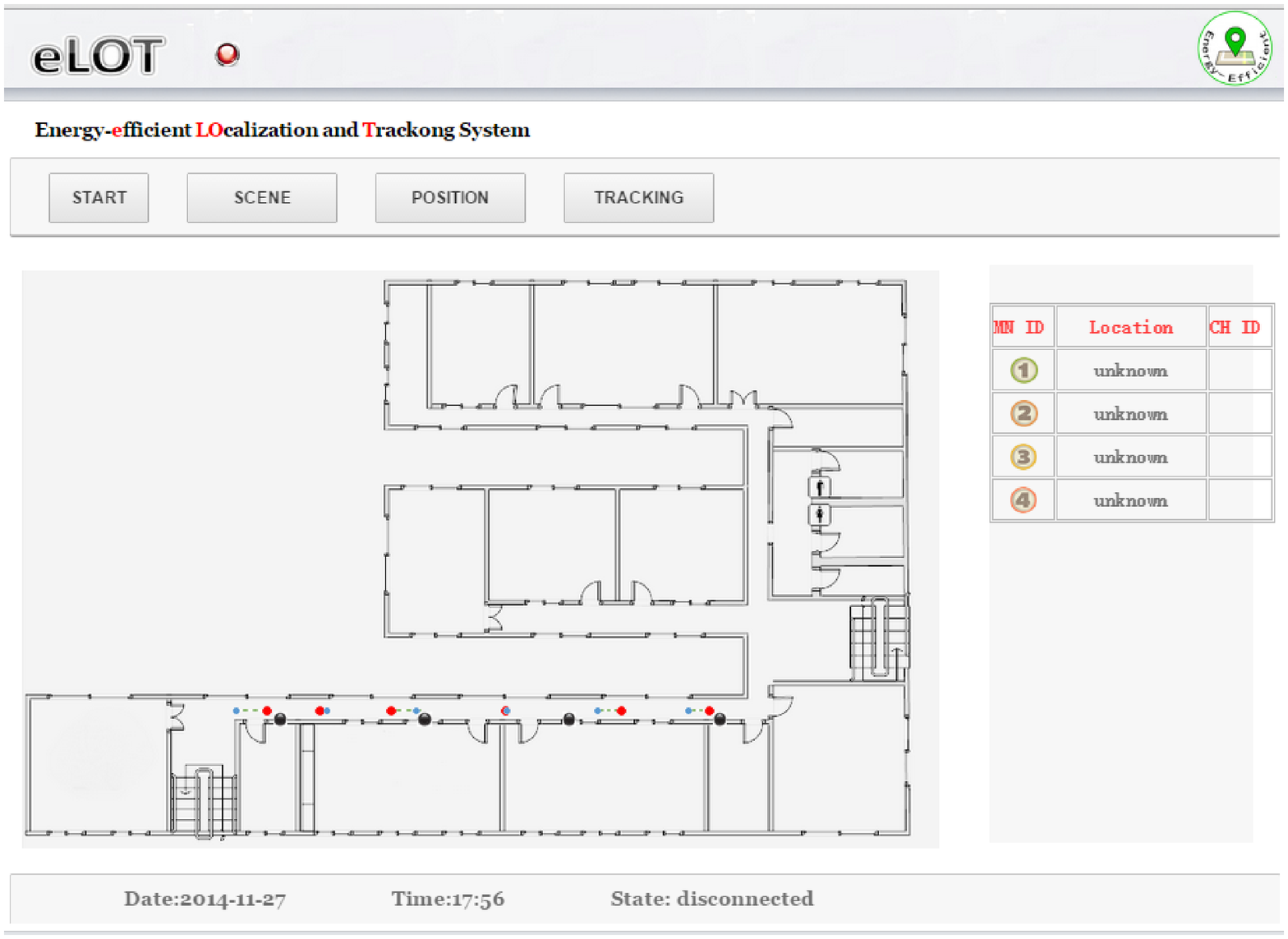}
\caption{An example web page in eLOT shown to the client.}
\label{alg_tracking}
\end{figure}

To order to enable the user to view the real-time position and to track the movement of the concerned MNs, the Web service module is programmed using the JavaScript language. The client browser first sends a connection request to the server using JavaScript. Then, the server approves it and establishes a TCP connection with the browser for data communications. On the web page, the positions of the fixed ANs are always marked in the map, while the real-time movement of the MNs can be observed under different scenarios. Apart from this, the user can click on the control bars to select functions such as scenarios, location and track. For example, when the location function is selected, only the user position is shown on the map without its movement routes as illustrated in Fig.~\ref{alg_tracking}.

\section{Experimental Results and Analysis}
\label{sec_Results}

\subsection{Experimental Setup}

\begin{figure}
\centering
\subfigure[Scenario A.]{\includegraphics[width=0.4\textwidth, angle=270]{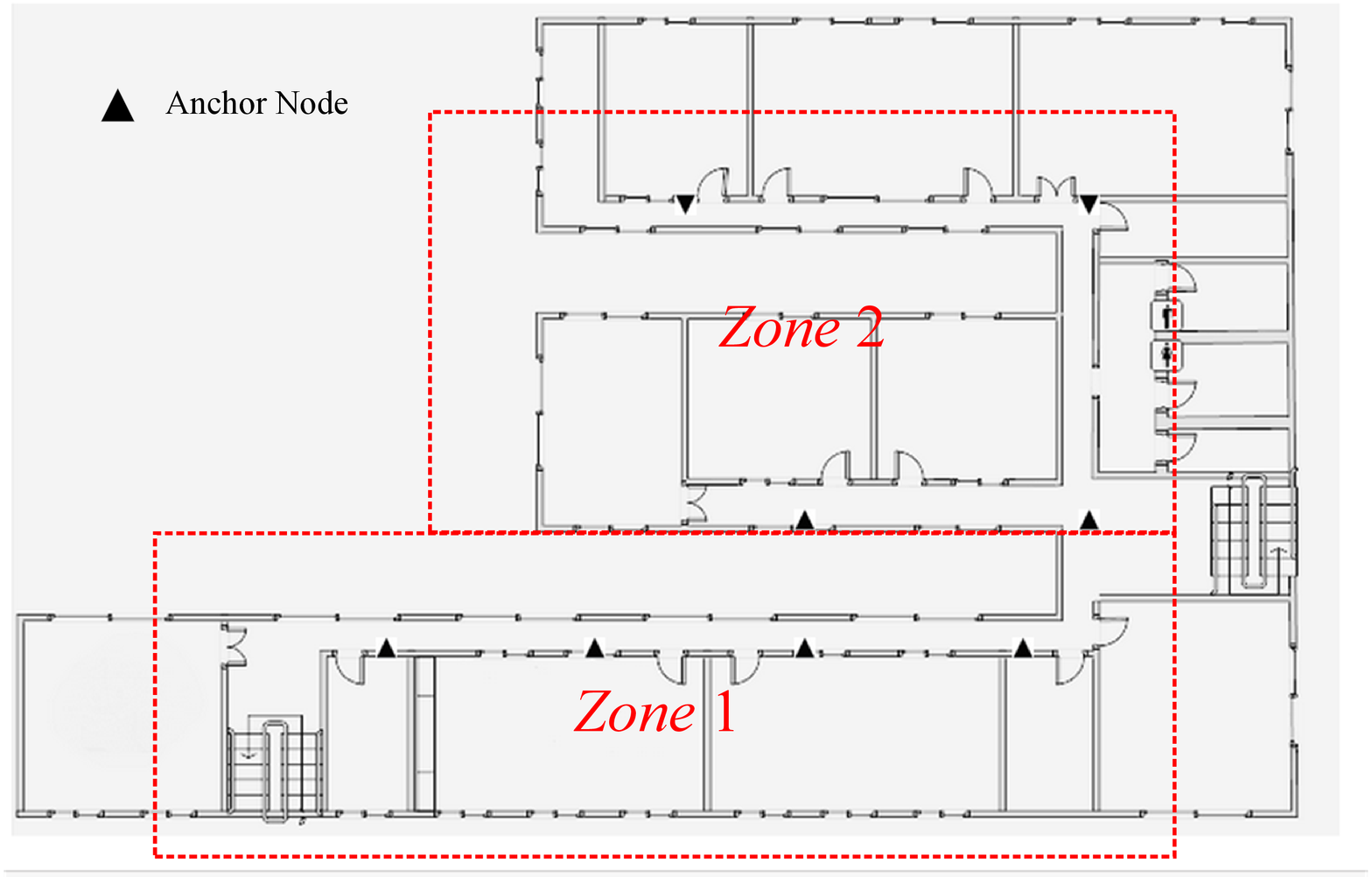}
\label{fig_SeNB_deploymentdeployment} }
\subfigure[Scenario B.]{\includegraphics[width=0.4\textwidth, angle=270]{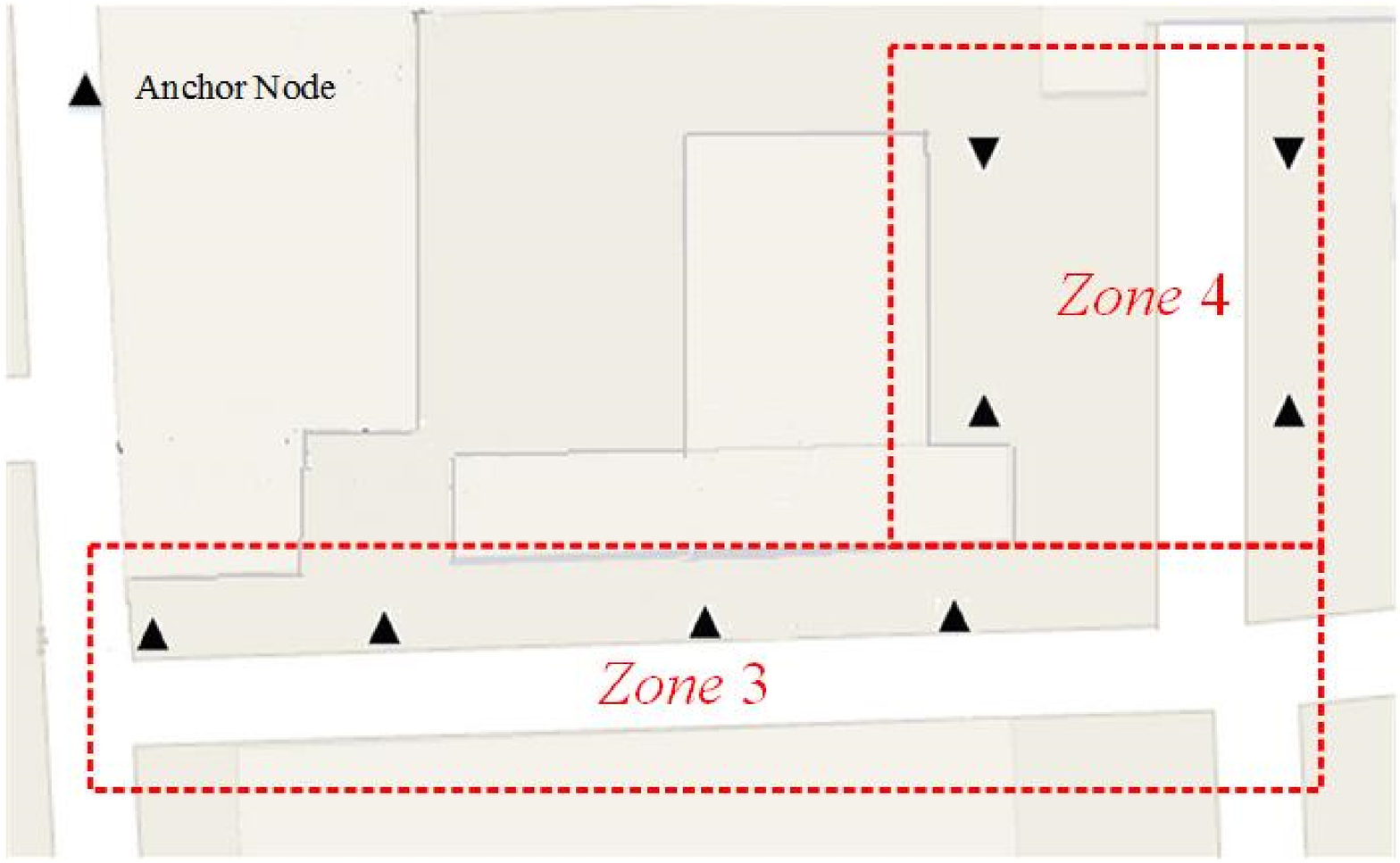}}
 \caption{Illustration of two typical deployment scenarios in eLOT.}
 \label{scenarios}
\end{figure}
%

We carry out experiments under two typical scenarios including both the indoor and outdoor environments. The first one (Scenario A) is a relatively small multi-floor apartment building, where each floor measures 200 square meters. As shown in Fig.~\ref{scenarios}(a),  the second floor is divided into two zones, i.e., \emph{Zone} 1 and \emph{Zone} 2. Due to geographical limitation, the areas of these two zones are not identical. We deploy four anchor nodes on the roof of the corridor in each zone, which results in an average of 8 meters separation in \emph{Zone} 1 and 10 meters separation in \emph{Zone} 2.  The wireless signals in this environment are relatively stable due to the small space.  The second scenario (Scenario B) is a typical Line-of-sight (LOS) outdoor environment as shown in Fig.~\ref{scenarios} (b). There are two zones in Scenario B, i.e., \emph{Zone} 3 and \emph{Zone} 4. In each zone, four anchor nodes are deployed along the road with an average separation of 15 meters, which are fixed on a tree or light pole approximately 2 meters above the ground. The experiments are conducted in an open field, without obstructions among the anchor nodes. \par

We collect fingerprints in different reference positions in Scenarios A and B. In Scenario A, 56 reference positions are picked, and the distance between two reference positions is 1.2 meters. In Scenario B, 12 reference positions with 2.4 meters separation are selected. Experiments are carried out when there are few people in Scenario B, causing little disturbance on the wireless signal. In both environments, we perform 30 signal sample scans for each reference position. Then, after removing the noise corrupted ones, the remaining samples are averaged and stored in the database as a fingerprint in the radio map. \par

Our chosen target is a person carrying a mobile node moving inside the detection area. In the online phase, we walk through the pre-defined routes multiple times. The 8-bit RSS indicator (RSSI) is used to represent the RSS value, ranging from -70 dBm and -10 dBm. The radio transmit power is set to 20 dBm. The RSS sampling time interval is adjustable in accordance with the target mobility. When the target velocity is below 0.5 $m/s$, the sampling interval is set to 1 $s$. The anchor nodes collect data from the mobile node. Measurements are sent to the server, i.e., Lenovo M4350, for data processing. Experiments in those environments require much time to complete. The accuracy of the eLOT system is found by comparing the estimated and known positions. \par

\subsection{Localization performances}

\begin{figure}
\centering
\includegraphics[width=0.5\textwidth, angle=270]{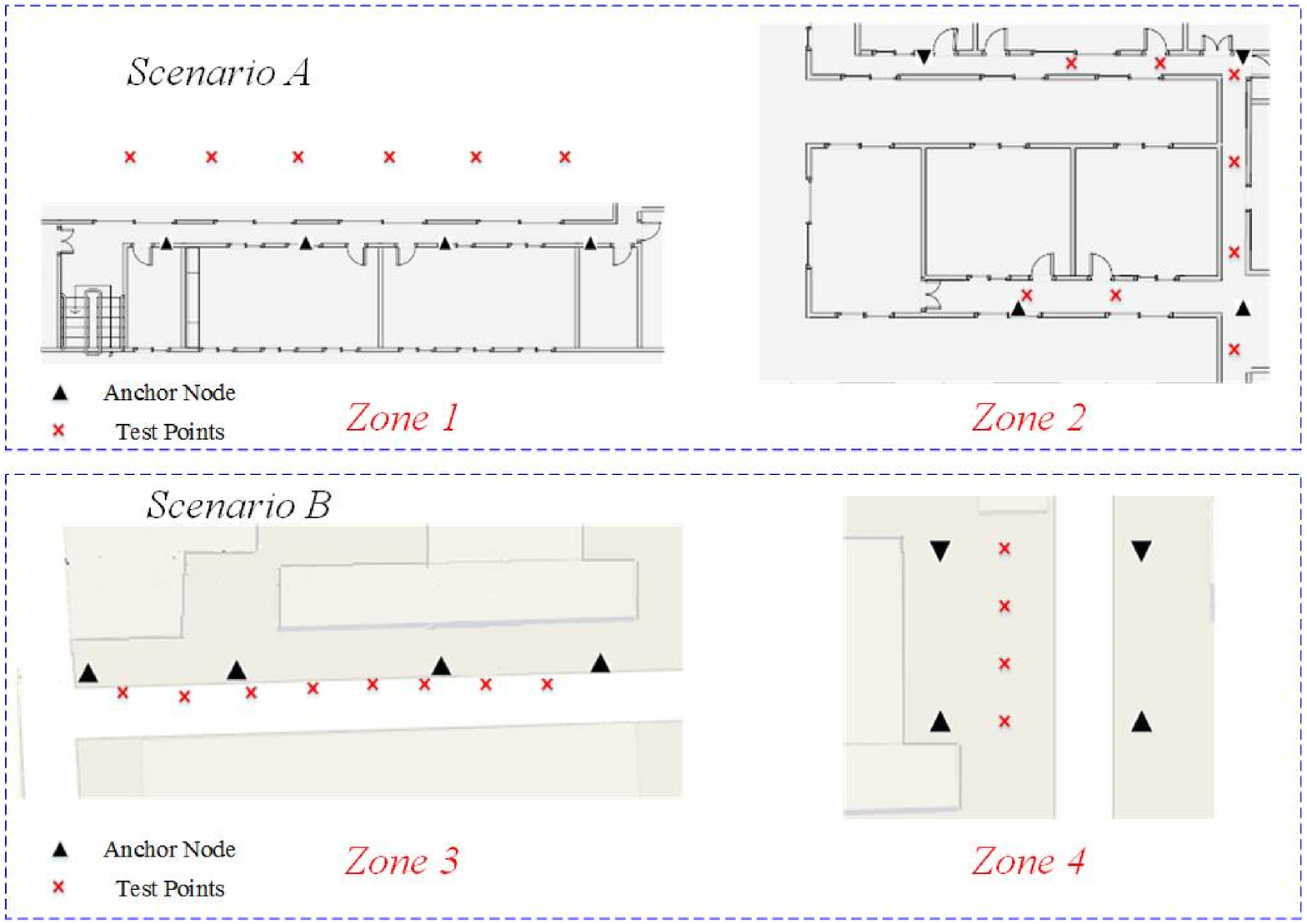}
\caption{Locations of the test points and reference positions.}
\label{TestLocation}
\end{figure}

In both environments, we perform tests at several given points as shown in Fig.~\ref{TestLocation}. At each points, more than 20 samples of the beacon signal from the testing MN are measured and used to estimate the position. \par

\begin{figure}
\centering
\includegraphics[width=0.6\textwidth]{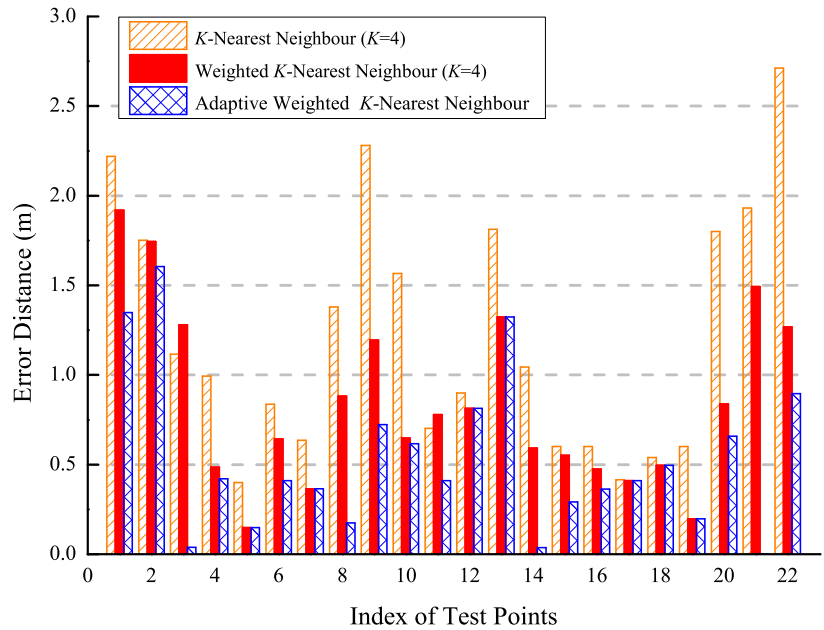}
\caption{Localization accuracy of different comparative algorithms for Scenario A.}
\label{fig_different_localization}
\end{figure}

The performances of different localization algorithms, i.e., KNN, WKNN and AWKNN, are compared for \emph{Zone} 1. Fig. \ref{fig_different_localization} illustrates the estimation errors of the comparative algorithms at the given point in \emph{Zone} 1. It is shown that the AWKNN algorithm outperforms its competitors, which is thus adopted in the following experiments. \par

\begin{figure}
\centering
\includegraphics[width=0.6\textwidth]{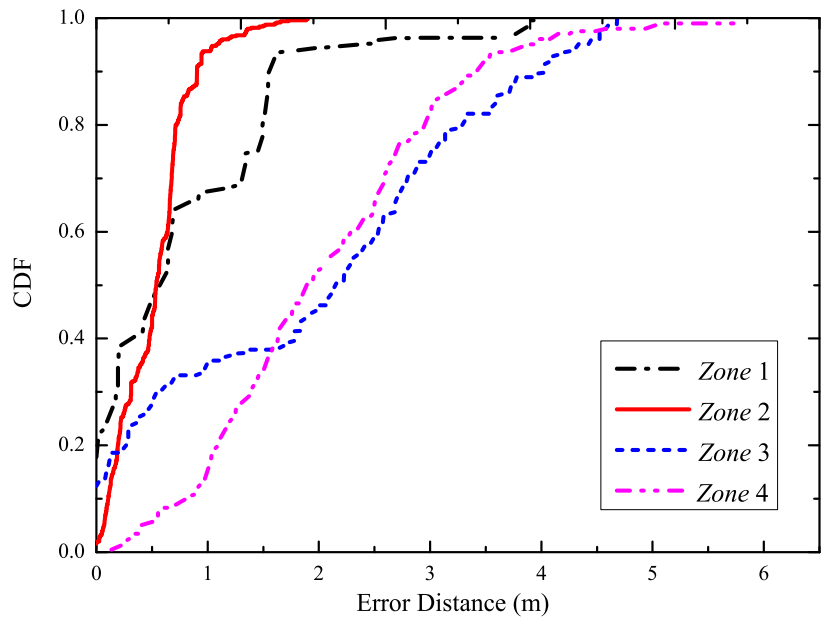}
\caption{CDF performances of the estimation errors in different zones.}
\label{fig_different_zone}
\end{figure}

Next, the cumulative distribution function (CDF) performances of the estimation errors under different zones are compared in Fig. \ref{fig_different_zone}. The average performances are also given in Table~\ref{errortable}.
Good localization accuracy is achieved for both the indoor and outdoor environments. In the indoor cases, i.e.,~\emph{Zone }1 and~\emph{Zone }2,  the accuracy is better due to the closer distance between the ANs, i.e., less than 2 $m$. On the other hand,  more errors were found for the outdoor cases, i.e.,~\emph{Zone} 3 and~\emph{Zone} 4, attributed to the more varying sounding environment and the larger distance between the ANs. \par

\begin{table}
\centering
\renewcommand{\arraystretch}{1.7}
\caption{Estimation errors in different zones. } \label{errortable} \centering
\begin{tabular}{|c|c|c|c|}
\hline
&\textbf{Average Error} & \textbf{ $<$ 50\% Error} &	\textbf{$<$  90\% Error} \\
 \hline
\emph{Zone} 1	&0.78 $m$	&0.56 $m$	&1.55 $m$ \\
 \hline
\emph{Zone} 2 	&0.53 $m$	&0.54 $m$	&0.90 $m$\\
 \hline
\emph{Zone} 3	&1.95 $m$	&2.14 $m$	&4.01 $m$\\
 \hline
\emph{Zone} 4	&2.03 $m$	&1.89 $m$	&3.38 $m$\\
 \hline
\end{tabular}
\end{table}

\subsection{Tracking performances}

\begin{figure}
\centering
\includegraphics[width=0.45\textwidth, angle=270]{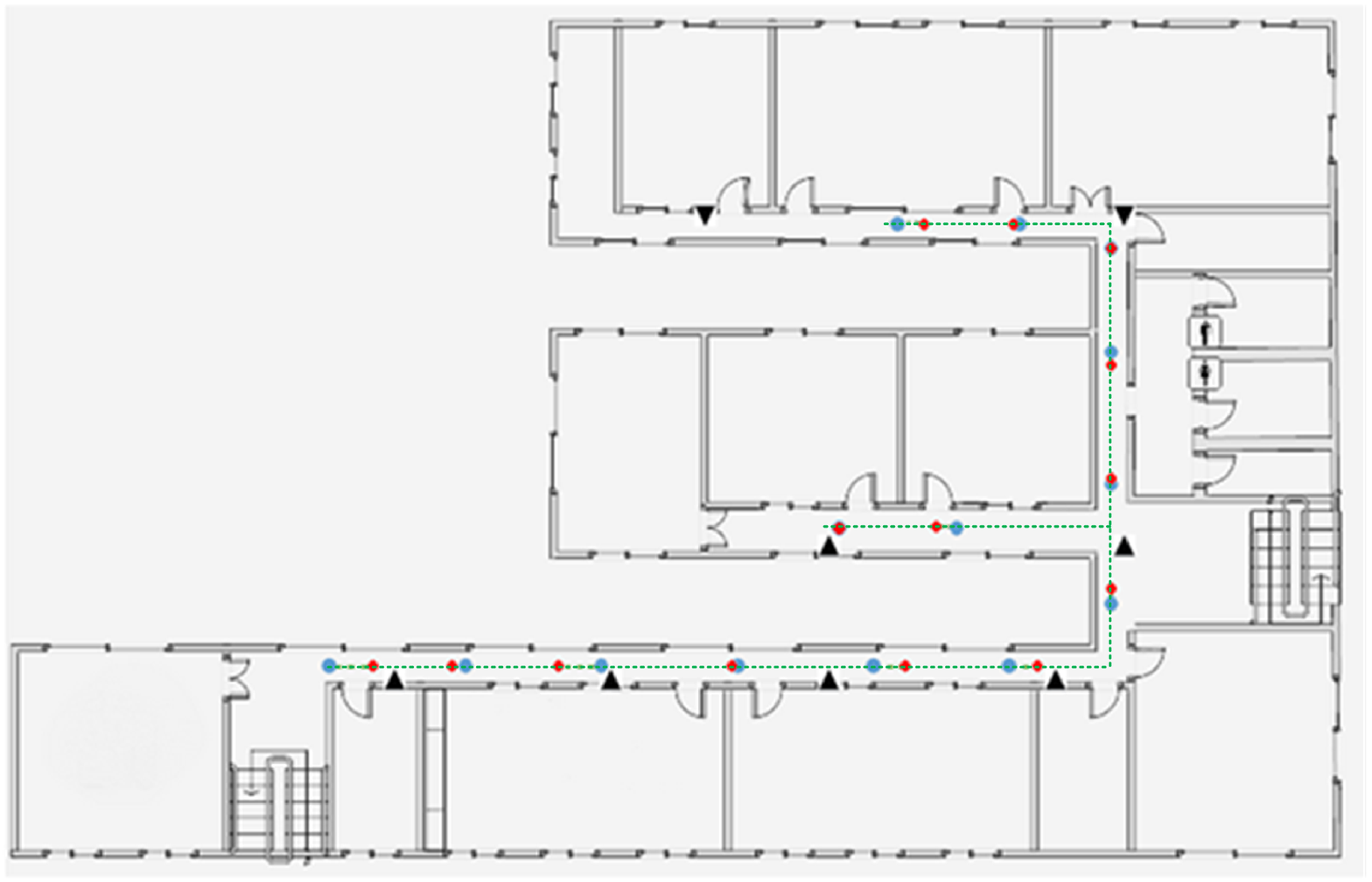}
\caption{Illustration of the tracking error in Scenario A.}
\label{fig_TrackError}
\end{figure}


In the online positioning phase, we walk through the pre-defined routes multiple times. The route
in Scenario A is represented by the dotted green lines in Fig.~\ref{fig_TrackError}, where the blue dots indicate the true target positions and the red ones represent the estimated positions. It can be seen that the errors are around one meter in this environment.

\subsection{Energy efficiency performances}

\begin{figure}
\centering
\includegraphics[width=0.6\textwidth,, angle=0]{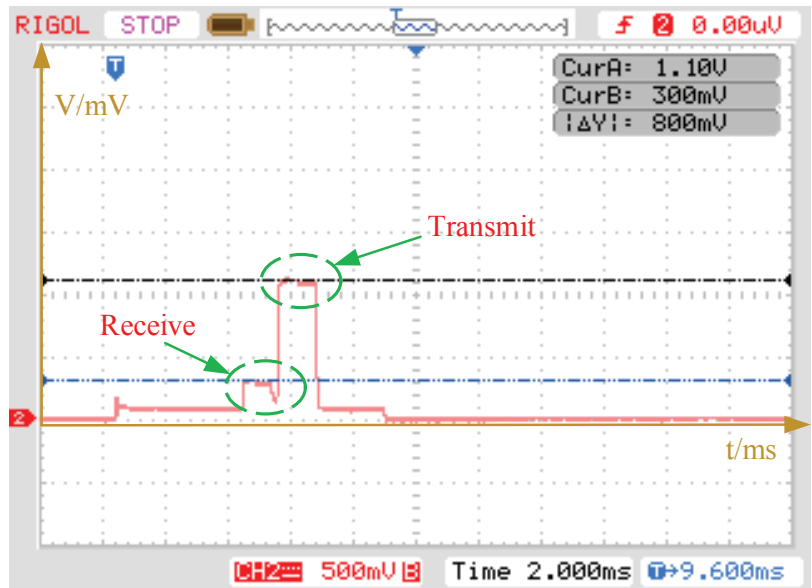}
\caption{Measured voltage results of the module in CC2530 and CC2591.}
\label{fig_voltage}
\end{figure}

In addition to position accuracy, energy consumption is also investigated in eLOT. First, the current consumption of the homegrown wireless communications module in CC2530 and CC2591 is measured by interpolating a series resistor of 10 $\Omega$. The measured voltage of the transmit or receive mode is shown in Fig.~\ref{fig_voltage}. Correspondingly, the currents of the transmit and receive mode are calculated, i.e., 110 $mA$ and 30 $mA$, respectively. Due to the limitation of the metering equipment, the current of the sleep mode is not measured but can be found in the data sheet, i.e., 1 $\mu A$~\cite{2530}. \par

\subsubsection{Effects of the proposed networking method}

In eLOT, more reliable communications can be achieved so that energy consumption decreases without unnecessary transmissions. In order to demonstrate its effectiveness in energy efficiency, tests were carried out for the case of multiple MNs working simultaneously in the multi-zones, i.e.,  two MNs in \emph{Zone} 1 and another two MNs in the neighbouring \emph{Zone} 2.  For the purpose of comparison, the performances of the traditional WSN without consideration of localization is also measured. All the MNs send the beacon signals to the ANs, and the successfully received packets are counted at the server. Thanks to the proposed networking methods, the packet loss rate (PLR) improves in eLOT compared with the traditional WSN, i.e., from  6.3\% reduced to 1.9\%.  With an increase in the number of MNs in the area, such improvements become more evident because the proposed networking methods can effectively reduce the competition among the MNs. \par



\subsubsection{Effects of adaptive sounding scheme}

\begin{figure}
\centering
\includegraphics[width=0.8\textwidth]{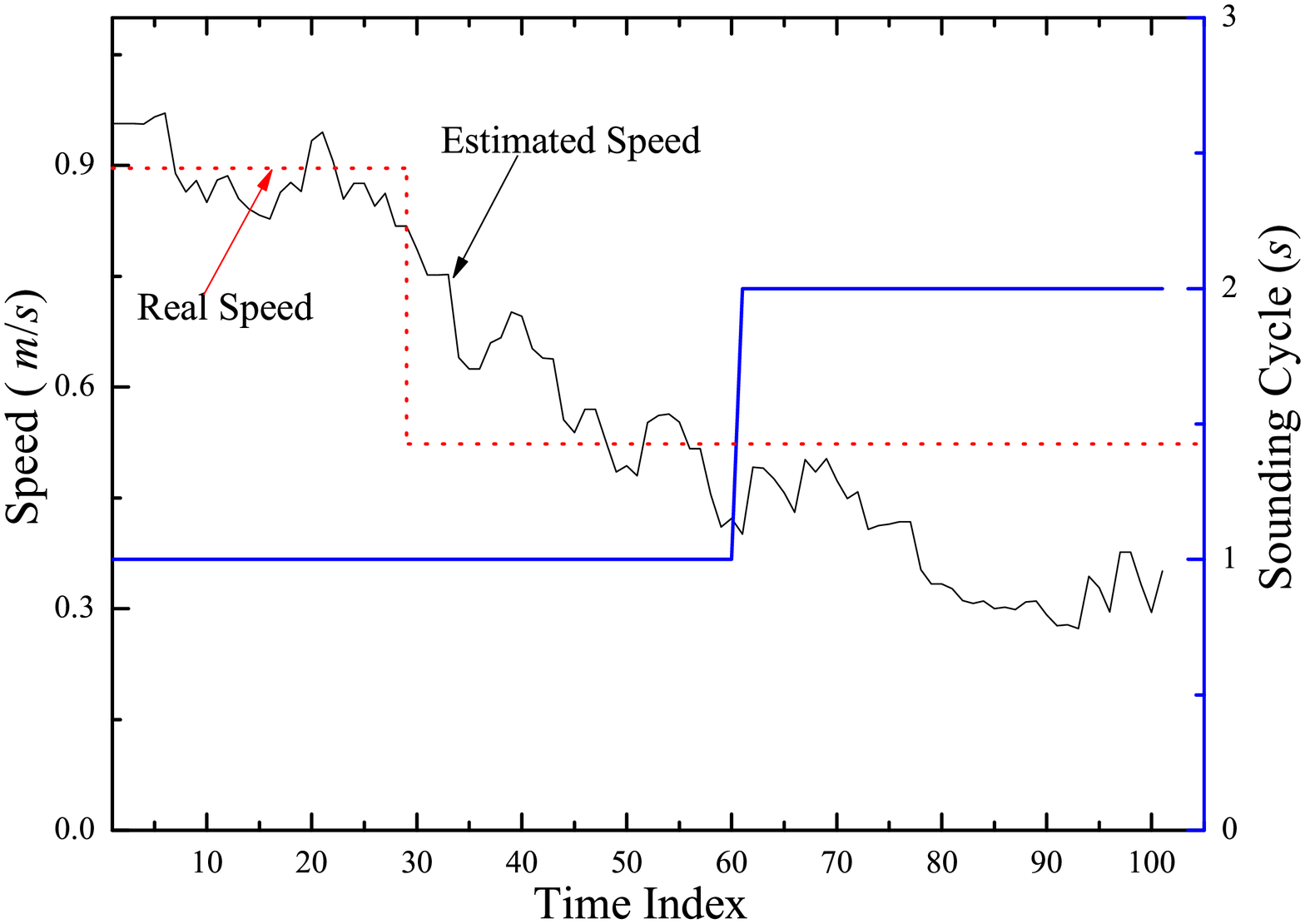}
\caption{Example of adjusting the sounding cycle during NM movement.}
\label{fig_SoundCycle}
\end{figure}

In the experiments to measure the effects of the adaptive sounding scheme, the mobility of the test MN is controlled, e.g., a speed of 0.9 $m/s$ is initialized at the beginning and reduced to around 0.5 $m/s$ later on.  As shown in Fig.~\ref{fig_SoundCycle}, the mobile node is real-time tracked although estimation errors exist. Then, the sounding cycle is adaptively adjusted according to the estimated speed, i.e., from 1 $s$ to 2 $s$. It is noted that such an adjustment does not take place immediately after the speed variance is estimated, which helps reduce the probability of erroneous changes. The energy consumption per sounding cycle, i.e., $T_{s}$, is given by
\begin{eqnarray}
\label{enegy}
E_{s}=P_{x} T_{x} +P_{s} (T_{s} - T_{x})
\end{eqnarray}
\noindent where  $P_{x}$ and $P_{s}$ are the power consumptions of the transmission mode and sleep mode, respectively, i.e., $3.3 \ V \times 110 \ mA$ and $3.3 \ V \times 1 \ \mu A$. $T_{x}$ is the time duration of the transmission mode, i.e., 1 $ms$.
Hence when the sounding cycle increases from 1 $s$ to 2 $s$, the energy consumption per sounding cycle increases very little, i.e., from 366 $\mu J $ to 370 $\mu J$, due to the very low energy consumption in the sleep mode.
In our test period of 100 $s$, the total energy consumption without the adaptive sounding scheme can be easily calculated as $100  \times 366 \mu J = 36600 \ \mu J =0.366 \ J $. On the other hand, with the adaptive scheme, the sounding cycle increases from 1 $s$ to 2 $s$ after detecting the speed change in the last 40 $s$, i.e., only 20 sounding cycles in 40 $s$. Correspondingly, the total energy consumption is $60  \times 366  \mu J + 20 \times 370 \mu J = 29360  \ \mu J = 0.2936 \ J$. The energy consumption is reduced from 0.366  $J$  to 0.2936 $ J$, which is equivalent to about 20\% improvement.  The improvement becomes lager if the test period becomes longer. Meanwhile, the average errors of localization with or without the adaptive sounding scheme are also measured as 1.43 $m$ and 1.39 $m$, respectively.\par

As defined in \cite{dominik}, the energy efficiency of localization in energy constrained sensor networks is given by
\begin{eqnarray}
\label{enegyefficiency}
\eta=\frac{1/e^{2}}{\varepsilon}
\end{eqnarray}
\noindent where $e^{2}$ is the Mean Square Error (MSE) of localization, and $\varepsilon$ is the energy consumed to localize a node. According to~\eqref{enegyefficiency}, the corresponding energy efficiency performance increases from 1.41 to 1.67 by adopting the adaptive sounding scheme in this case. \par

\section{Conclusion}
\label{Conclusions}
We developed a Zigbee-based localization and tracking system, which is applicable in ubiquitous environments including both indoor and outdoor. In order to well balance localization accuracy and energy efficiency, new designs at both the network and mobile nodes were implemented in the proposed system, where the mobile nodes can have quick access to the network and be located by the network with minimum energy consumption. Extensive experiments were conducted in practical environments to evaluate the performance of our system. The results have demonstrated that eLOT is energy-efficient and effective in accurately estimating target positions in various setting. In our future work, we will enlarge the network size of eLOT to include multiple buildings as well as roads in the campus. \par



\begin{thebibliography}{99}


\bibitem{ref1}
D.~Cai, ``A retail application based on indoor location with grid estimations," in \textit{Proc. International Conference on
Computer, Information and Telecommunication Systems (CITS)'2014}, Jeju, 7-9 July, 2014.

\bibitem{ref2}
S.~ Chen, Y.~Chen and  W.~Trappe, ``Inverting Systems of Embedded Sensors for Position Verification in Location-Aware Applications," in \textit{IEEE Transactions on Parallel and Distributed Systems}, vol. 21, no. 5, pp. 722 -736, 2010. 	







\bibitem{Cenedese_TVT_10}
A.~Cenedese, G.~ Ortolan, and M.~Bertinato , ``Low-Density Wireless Sensor Networks for Localization and Tracking in Critical Environments,"" \textit{IEEE Transactions on Vehicular Technology}, vol. 59, no. 6, pp. 2951 - 2962, 2010.

\bibitem{Zhou_TVT_10}
J.~Zhou; J.~Shi and X. Qu, ``Landmark Placement for Wireless Localization in Rectangular-Shaped Industrial Facilities," \textit{IEEE Transactions on Vehicular Technology}, vol. 59, no.6, pp. 3081 -3090, 2010.






\bibitem{KNN}
A.~Kokkinis, etl,, ``Map-aided fingerprint-based indoor positioning," in \textit{Proc. IEEE Personal Indoor and Mobile Radio Communications (PIMRC)'2013 }, London, United Kingdom, Sept., 2013.

\bibitem{MMSE}
R.~Teemu, etl, ``A probabilistic approach to WLAN user location estimation.
\textit{International Journal of Wireless Information Networks}, vol.9, no.3, pp. 155 - 164,
July 2002.

\bibitem{K03}
K.~Kleisouris, Y.~ Chen, J.~ Yang, and R.~Martin, ``Empirical evaluation of wireless localization when using multiple antennas," \textit{IEEE Transactions on Parallel and Distributed Systems}, vol. 21, no. 11, pp. 1595-1610, 2010.	

\bibitem{OpenMAC}
A. V. Medina, J. A. Gomez, O. Rivera, and E. Dorronzoro, ``Fingerprint indoor position system based on OpenMAC," in \textit{Proc. International Conference on Wireless Information Networks and Systems (WINSYS)},Seville, Spain, 18-21 July, 2011.



\bibitem{NN}
G. Chen, Y. Zhang, L. Xiao, J Li, L Zhou, and S. Zhou, ``Measurement-based RSS-multipath neural network indoor positioning technique," in \textit{Proc. IEEE Canadian Conference on Electrical and Computer Engineering (CCECE)'2014}, Toronto, ON, 4-7 May, 2014.


\bibitem{Arya09}
A. Rya, and P. Godlewski, ``Performance analysis of outdoor localization systems based on RSS fingerprinting," in \textit{Proc. International Symposium on Wireless Communication Systems} Tuscany, 7-10 Sept., 2009.

\bibitem{smart}
Y. Kim, Y. Chon, and H. Cha, ``Smartphone-Based collaborative and autonomous radio fingerprinting," \textit{IEEE Transactions on Systems, Man, and Cybernetics, Part C: Applications and Reviews}, vol. 42 , no. 1, pp. 112 - 122, 2012.




\bibitem{GZ08}
G. Zanca, F. Zorzi, A. Zanella, and M. Zorzi, ``Experimental comparison
of RSSI-based localization algorithms for indoor wireless sensor
networks," in \textit{Proc. REALWSN'2008}, pp. 1 - 5, April 1, 2008.


\bibitem{radioerror}
M. Pichler, S. Schwarzer, A. Stelzer, and M. Vossiek, `` Multi-Channel distance measurement with IEEE 802.15.4 (ZigBee) devices,"
\textit{IEEE Journal of Selected Topics in Signal Processing}, vol. 3, no. 5, pp. 845 -859, 2009.




\bibitem{PTL}
J. Graefenstein and M. E. Bouzouraa, ``Robust method for outdoor
localization of a mobile robot using received signal strength in
low power wireless networks," in \textit{Proc. IEEE ICRA 2008}, May, 2008.

\bibitem{it}
E.  Goldoni, and A.~Savioli, M.~Risi, and P.~Gamba
, ``Experimental analysis of RSSI-based indoor localization with IEEE 802.15.4," in \textit{Proc. European
Wireless Conference (EW)'2010}, Lucca, 12-15 April, 2010.



%
\bibitem{tvt_coop}
T.~Van Nguyen, Y.~ Jeong, H.~ Shin and M.~Win, ``Least Square Cooperative Localization," \textit{IEEE Transactions on Vehicular Technology}, vol.64, no. 4, pp. 1318 -1330, 2015.

\bibitem{pd_coop}
W.~ Li, Y.~Hu, X.~ Fu, S.~Lu, and D.~Chen, ``Cooperative Positioning and Tracking in Disruption Tolerant Networks," \textit{IEEE Transactions on Mobile Computing}, vol. 26, no. 2, pp. 382- 391, 2015.


\bibitem{tmc_deployment}
M.~Ficco, C.~ Esposito, and A.~ Napolitano, ``Calibrating Indoor Positioning Systems
with Low Efforts," \textit{IEEE Transactions on Mobile Computing}, vol. 13, no. 4, pp. 737 - 751, 2014.


\bibitem{EEsurvey}
G. Anastasi, M. Conti, M.~Francesco, and A. Passarella, ``Energy conservation in wireless sensor networks: a survey,"
 \textit{Ad Hoc Networks}, vol.7, No. 3, May 2009.


\bibitem{Cheng13}
P.~Cheng, F.~Zhang, J.~Chen, Y.~Sun and X.~Shen, ``A Distributed TDMA scheduling algorithm for target tracking in ultrasonic sensor networks," \emph{IEEE Transactions on Industrial Electronics}, vol. 60, no.9, pp. 3836-3845, 2013.


\bibitem{Shu15}
Y.~Shu, P.~Cheng, Y.~Gu, J.~Chen, and T.~He, ``TOC: Localizing wireless rechargeable sensors with time of charge," \emph{ACM Transactions on Sensor Networks}, vol. 11, no. 3, Article 44, May, 2015.




\bibitem{EEtrackingsurvey}
O. Demigha, W. Hidouci, and T. Ahmed, ``On energy efficiency in collaborative target tracking in wireless sensor network: a review,"
 \textit{IEEE Communications Surveys \& Tutorials}, vol. 15, no. 3, pp. 1210 - 1222, May, 2012.


\bibitem{tmc4}
M. Ficco, C. Esposito, and A. Napolitano, ``Toward Accurate Mobile Sensor Network Localization in Noisy Environments," \textit{IEEE Transactions on Mobile Computing}, vol. 12, no. 6, pp. 1094 -1106, 2013.

\bibitem{ZiFIND}
Y.~Xiong, R.~Zhou, M.~Li, G.~Xing, L.~ Sun, and J.~ Ma, ``ZiFind: exploiting cross-technology interference signatures for wireless LAN discovery," \emph{IEEE Transactions on Mobile Computing}, vol.13, no. 11, pp. 2675 - 2688, Nov. 2014.


\bibitem{ville}
V. Kaseva, T.D. Hamalainen,and M. Hannikainen, ``Range-free algorithm for energy-efficient indoor localization in wireless sensor networks," in \textit{Proc. Conference on Design and Architectures for Signal and Image Processing (DASIP)'2011}, pp. 1 -8, Nov., 2011.


\bibitem{tmc3}
F. Barsi, A. Bertossi, and C. Lavault, ``Efficient Location Training Protocols for Heterogeneous Sensor and Actor Networks," \textit{IEEE Transactions on Mobile Computing}, vol. 10, no. 3, pp. 377- 391, 2011.

\bibitem{coal}
Y.~Liu, S.~ Lu and Y.~Liu, ``COAL: Context aware localization for high energy efficiency in wireless networks,"  in \textit{Proc. IEEE Wireless Communications and Networking Conference (WCNC)'2011}, pp. 2030 - 2035, March, 2011.

\bibitem{santosh}
S. Pandey, P. Prasad, P.~Sinha, and P.~Agrawal, ``Localization of sensor networks considering energy accuracy tradeoffs," in \textit{Proc. International Conference on Collaborative Computing: Networking, Applications and Worksharing'2005}, Dec. 2005.

\bibitem{dominik}
D.~Lieckfeldt, J.~ You, R.~ Behnke, J.~ Salzmann, and D.~Timmermann, ``Assessing the energy efficiency of localization in wireless sensor networks," in \textit{Proc. IEEE Consumer Communications and Networking Conference (CCNC)'2009}, pp.1 -2, Jan. 2009.


\bibitem{2530}
CC253x User's Guide : CC253x System-on-Chip Solution for 2.4 GHz IEEE 802.15.4 and ZigBee
Applications (SWRU191).

\bibitem{802154}
IEEE Std. 802.15.4-2006: Wireless Medium Access Control (MAC) and Physical Layer (PHY) Specifications or Low-Rate Wireless Personal Area Networks (LR-WPANs)
http://standards.ieee.org/getieee802/download/802.15.4-2006.pdf

\end{thebibliography}
\end{document}